\documentclass[cits]{PoS2}
\usepackage{amsmath}
\usepackage{subfigure}

\def\tapp{\theta_{\rm app}}
\def\lesssim{\mathrel{\hbox{\rlap{\hbox{\lower4pt\hbox{$\sim$}}}\hbox{$<$}}}}
\def\gtrsim{\mathrel{\hbox{\rlap{\hbox{\lower4pt\hbox{$\sim$}}}\hbox{$>$}}}}

\title{AGN jet physics and apparent opening angles}

\ShortTitle{AGN jet physics and apparent opening angles}

\author{\speaker{Eric Clausen-Brown} $^{1}$\thanks{clausenbrown@mpifr.de}, Tuomas Savolainen$^1$, Alexander B. Pushkarev$^{2,3,1}$, Yuri Y. Kovalev$^{4,1}$, Matthew L. Lister$^5$\\
\llap{$^1$}Max-Planck-Institut f\"{u}r Radioastronomie, Auf dem H\"{u}gel 69, 53121 Bonn, Germany\\
\llap{$^2$}Pulkovo Astronomical Observatory, Pulkovskoe Chaussee 65/1, 196140 St. Petersburg, Russia\\
\llap{$^3$}Crimean Astrophysical Observatory, 98409 Nauchny, Crimea, Ukraine\\
\llap{$^4$}Astro Space Center of Lebedev Physical Institute, Profsoyuznaya 84/32, 117997 Moscow, Russia\\
\llap{$^5$}Department of Physics, Purdue University, 525 Northwestern Avenue, West Lafayette, IN 47907, USA}

\abstract{
We present a new method to measure $\Gamma \theta_j$ in flux-limited samples of active galactic nuclei (AGN) jets, where $\Gamma$ is the bulk Lorentz factor and $\theta_j$ is the jet's half-opening angle.  The $\Gamma \theta_j$ parameter is physically important for models of jet launching, and also determines the effectiveness of jet instabilities and magnetic reconnection.  We measure $\Gamma\theta_j$ by analyzing the observed distribution of apparent opening angles in very long baseline interferometry (VLBI) flux-limited samples of jets, given some prior knowledge of the active galactic nuclei (AGN) radio luminosity function.  
We then apply this method to the MOJAVE flux-limited sample of radio loud objects and find $\Gamma \theta_j \approx 0.1\pm 0.03$, which implies that AGN jets are subject to a variety of physical processes that require causal connection.  }

\FullConference{11th European VLBI Network Symposium \& Users Meeting,\\
		October 9-12, 2012\\
		Bordeaux, France}

\begin{document}
\section{Introduction}
A relativistic jet's value of the parameter $\Gamma\theta_j$ can impact jet physics for several different reasons, where $\Gamma$ is the bulk Lorentz factor and $\theta_j$ is the jet's half-opening angle.  In some jet models, $\Gamma\theta_j$ is expected to be $\sim 1$ if either the jet's Poynting flux is efficiently converted to kinetic flux \cite{Komissarov2009}, or if the jet is freely expanding.   For hydrodynamic models, $\Gamma\theta_j$ is related to the ratio of the pressure of the ambient medium to the jet's internal pressure \cite{Daly1988}.  Also, most jet acceleration models require that the jet be in causal contact, which implies that $\Gamma\theta_j<1$.  Besides global jet dynamics, $\Gamma\theta_j$ may also have significance for transient events such as magnetic reconnection and jet instabilities, since the jet must be in causal contact for such processes to occur \cite{Begelman1998}, and $\Gamma\theta_j$ in part controls how much jet sidewise expansion slows instability growth \cite{Giannios2006,Narayan2009} .  Additionally, $\Gamma\theta_j$ is a significant parameter affecting the structure of shocked boundary layers \cite{Kohler2012a} and the efficiency of particle acceleration at recollimation shocks \cite{Nalewajko2009}.

There have been two past measurements of the characteristic value of $\Gamma\theta_j$ for AGN jets.  Using 7mm very long baseline array (VLBA) data from 15 different AGN jets, Jorstad et al.~\cite{Jorstad2005} measured $\Gamma\theta_j$ by assuming that the observed pattern speed of moving jet components corresponds to the jet bulk flow speed, and that the component variability times are equal to the jet frame light crossing times of the (resolved) components.  From these assumptions they determined the component's Lorentz factor and viewing angle, and found an anti-correlation between the derived values of $\theta_j$ and $\Gamma$, with $\Gamma\theta_j=0.17 \pm 0.08$.  With a larger sample of 56 AGN jets from 15 GHz VLBA data, Pushkarev et al.~\cite{Pushkarev2009} performed the same analysis, except they used Lorentz factors and viewing angles from Hovatta et al.~\cite{Hovatta2009}, who derived these values used variability time, maximum flux density of flares, and equipartition derived brightness temperature arguments.  Pushkarev et al.'s \cite{Pushkarev2009} analysis found a similar anti-correlation with $\Gamma\theta_j=0.13$.

Motivated by the above theoretical concerns, we set out here to provide a new method of measuring $\Gamma\theta_j$ by using a larger sample of AGN jets and very different physical assumptions.  We do this by deriving the expected distribution of apparent opening angles in a flux limited sample with $\Gamma\theta_j$ as a free parameter to be fixed by finding the best fit to an empirical distribution of apparent opening angles.  Our data consist of the 135 AGN jets that make up the MOJAVE-I sample, a 15 GHz flux density-limited survey conducted by the VLBA of radio sources in the northern sky with flux densities above 1.5 Jy, and above 2 Jy for sources with $-20<$ dec $<0$ \cite{Lister2009}.

\section{Constraining AGN jet physics with observed apparent opening angles}
Here we derive the probability density function (PDF) for apparent opening angles, $P(\tapp)$, and then fit it to a distribution of apparent half-opening angles, $\tapp$, measured from MOJAVE-I data.  To test the viability of the simplest case scenario, we assume that $\Gamma\theta_j$ is constant for all relativistic AGN jets.  As we show below, $\Gamma\theta_j$ is then a free parameter of $P(\tapp)$, and thus will be determined in the fit to the empirical distribution of $\tapp$.

To derive $P(\tapp)$, we first derive the PDF for viewing angles, $P(\theta)$.  If the sample of AGN jets were unbiased with respect to orientation, $P(\theta)=\sin{\theta}$.  However, because it is flux-limited, it will take the form
\begin{equation}
P(\theta)=\mbox{\emph{Doppler bias factor}} \times \sin{\theta}.  
\end{equation}
This additional factor takes into account that more sources are directed at the observer in a flux limited sample because Doppler beamed jets are detectable at further distances than unbeamed ones.  Vermeulen \& Cohen \cite{Vermeulen1994} computed this term, finding that it depends on the bulk Lorentz factor distribution in a flux limited sample (every jet is assumed to possess a single Lorentz factor), the integral source count index, and the beaming index of the jet.  The beaming index is defined from the relation $F=\delta^nF'$, where $n$ is the beaming index, $\delta$ is the Doppler factor, $F$ is the observed flux density, and $F'$ is the intrinsic flux density.  The observed integral source count index is defined in the expression $N(>F)\propto F^{-q}$, representing the number of sources $N$ observed with a flux density above $F$, which is a power law in $F$ with source count index $q$.  Including the Doppler bias factor in the viewing angle PDF gives
 \begin{equation}
 P(\theta,\Gamma) =A \left(1-\beta\cos{\theta}\right)^{-a-1}\sin{\theta}P(\Gamma),
 \label{VC}
 \end{equation}
where $a=nq-1$, $P(\Gamma)$ is the PDF for jet bulk Lorentz factor, and $A$ is the constant of normalization.  An important assumption made in calculating the Doppler bias term is that the log-log slope of $N(>F')$ vs $F'$ and $N(>F)$ vs. $F$ are the same, which Vermeulen \& Cohen \cite{Vermeulen1994} justify based on previous studies of AGN jet luminosity functions.

The opening angle distribution may now be derived from $P(\theta,\Gamma)$ by a change of variables from $\theta$ to $\tapp$, and marginalizing over $\Gamma$:
\begin{equation}
P(\tapp)=\int{d\Gamma P\left(\theta(\tapp,\Gamma),\Gamma\right)\left| \frac{\partial \theta}{\quad\partial \tapp} \right|},
\label{ptapp}
\end{equation}
where the $\theta$ and $\partial \theta/\partial \tapp$ are functions of $\tapp$ and $\Gamma$, and can be determined by assuming a particular jet geometry, which we take to be conical here.  These relationships are often derived by treating conical jets as triangles projected onto the plane of the sky, implying that $\tan{\theta_{\rm app}}=R_j/\ell'=R_j/(\ell\sin{\theta})$, where $\theta$ is the jet viewing angle, $R_j$ is the jet radius, $\ell$ is the jet length, and $\ell'$ is the jet length projected onto the sky.  If we assume $\theta_j\approx R_j/\ell$, then
\begin{equation}
\tan{\theta_j}=\tan{\tapp}\sin{\theta}.
\label{tri}
\end{equation}
For $\tapp,\theta_j \ll 1$, this reduces to the most commonly used relation for jets, $\theta_j=\tapp \sin{\theta}$ \cite{Jorstad2005,Pushkarev2009}.  However, this relation breaks down as $\theta\rightarrow\theta_j$, since when $\theta=\theta_j$, equation (\ref{tri}) incorrectly gives $\theta_{\rm app}=\pi/4$ (it should be $\pi/2$ in this case because the projected jet streamlines would cover half of all directions on the sky).  However, we ignore such situations here because most jets are viewed such that $\theta>\theta_j$ \cite{Pushkarev2011}.

We now calculate an approximate analytic expression for $P(\tapp)$ and show that it is not very sensitive to $P(\Gamma)$.  Note that in a flux-limited VLBI sample, jets with small viewing angles will dominate, so we assume $\theta\ll 1$ and $\Gamma\gg 1$, and approximate (\ref{VC}) as
\begin{equation}
P(\theta,\Gamma)=A \frac{1}{(2\Gamma^2)^{a+1}}(1+\Gamma^2\theta^2)^{-a-1}\theta.
\label{approx}
\end{equation}
For simplicity, we now evaluate equation (\ref{ptapp}) in light of the geometry implied by equation (\ref{tri}), with the additional approximation that $\sin{\theta}\approx \theta$, and obtain
\begin{align}
P(\tapp)&=A\left(1+\frac{\rho^2}{\tan^2{\tapp}}\right)^{-a-1}\frac{\cos{\tapp}}{\sin^3{\tapp}}\left[\int{\frac{P(\Gamma)}{(2\Gamma^2)^{a+1}}d\Gamma}\right] \notag\\
&= A'\left(1+\frac{\rho^2}{\tan^2{\tapp}}\right)^{-a-1}\frac{\cos{\tapp}}{\sin^3{\tapp}}
\label{approx2}
\end{align}
where $\rho=\Gamma\theta_j$, and we have absorbed the term in brackets into the new normalization, $A'$.  Thus, it is apparent from equation (\ref{approx2}) that $P(\tapp)$ does not depend significantly on the form of $P(\Gamma)$.  Equation (\ref{approx2}) is an accurate approximation of $P(\tapp)$ as long as $\rho\ll1$, which, as shown below, is a valid assumption.

\begin{figure}[t]
\centering
\includegraphics[width=4.0in]{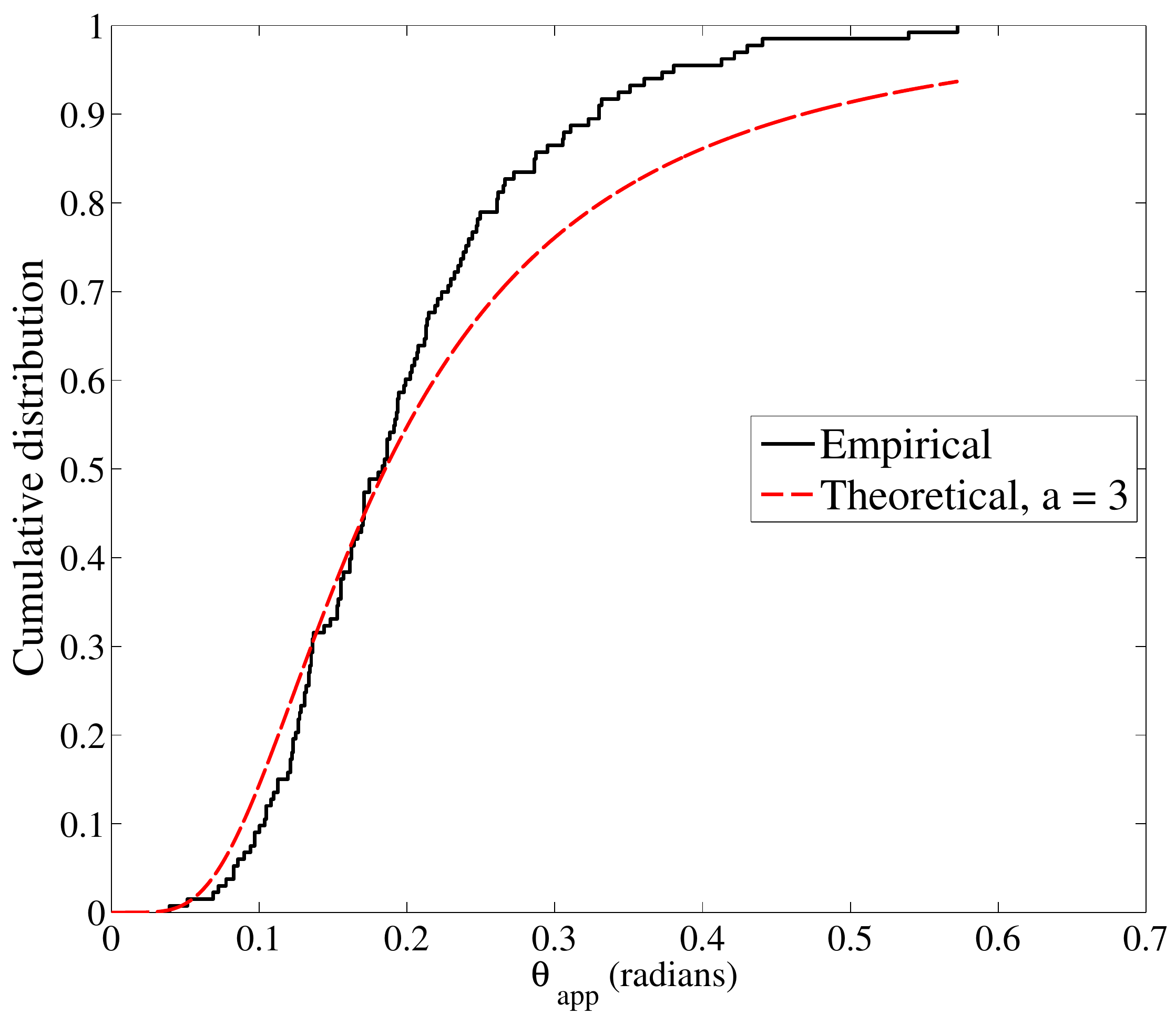}
\caption{Fitting apparent opening angle data to the derived probability density (eqn. \protect\ref{approx2}) via maximum likelihood estimation.  The only free parameter is $\rho$, and $a=3$ is assumed.  The result is $\rho^{\rm fit}=0.095 \pm 0.009$.  Note that the best fit cumulative distribution function (CDF) is systematically lower than the empirical CDF for $\tapp\gtrsim 0.2$.  }  
\label{datafit}
\end{figure}

We now fit equation (\ref{approx2}) to the empirical opening angle distribution from the MOJAVE-I data set using maximum likelihood estimation (MLE),  as shown in figure \ref{datafit}.  We set $a=3$, so our fits only include one free parameter, $\rho$ ($=\Gamma \theta_j$).  This value of $a$ is close to the value it would have if the beaming index corresponded to a steady jet \cite{Lind1985} with a spectral index of $\alpha=0.7$ ($F_{\nu}\propto \nu^{-\alpha}$) \cite{Pushkarev2012}, and $q=1.5$ \cite{Vermeulen1994}.  The result we obtain is $\rho_{\rm fit} = 0.095 \pm 0.009$ using equation (\ref{tri}).  The confidence interval, calculated from the inverse of the information matrix, is not very meaningful given the uncertainty in $a$ discussed below and other theoretical uncertainties discussed in \S\ref{conclusion}. 

The true confidence interval in the best fit value of $\rho$ requires analyzing $a$, whose theoretically likely values range from $\sim 2$ to $5$.  This range originates from beaming models of jets, which bound the beaming index $n$ as being between $2+\alpha$ and $3+\alpha$, and the measured integral source count index of $q\approx 1.5$ \cite{Vermeulen1994}.  The spectral index, $\alpha$, for jets is typically between  $0$ and $\sim 1$ \cite{Pushkarev2012}, thus given the range of $n$ as $2+0$ to $3+1$, we find the range of $a$ as being between $2\times1.5-1=2$ and $4\times1.5-1=5$.  Assuming such a range in $a$ alone determines the confidence interval of the best fit, then $\rho_{\rm fit}=0.096\pm 0.025$. 

To better understand how best fit values of $\rho$ depend on the assumed value of $a$, we calculate the best fit value of $\rho$ for different given values of $a$.  We find that the resulting curve of $\rho_{\rm fit}$ versus $a$ is very well reproduced by assuming that each best fit value of $\rho$ is sensitive to the location of the peak of the empirical number density distribution, $\tapp^{\rm peak}\sim 0.1$ to $0.2$, which is the inflection point of the cumulative distribution shown in figure \ref{datafit}.  This assumption implies that the relationship between $\rho$ and $a$ is can be found by maximizing the probability density $P(\tapp)$, which according to equation (\ref{approx2}) gives $\rho\approx \tapp^{\rm peak}\sqrt{3}/\sqrt{2a-1}$, or
\begin{equation}
\Gamma\theta_j\approx\frac{\tapp^{\rm peak}\sqrt{3}}{\sqrt{2(nq-1)-1}}.
\label{estimate}
\end{equation}
For an empirical peak of $\tapp^{\rm peak}=0.13$, this correctly reproduces the best fit values of $\rho$ for values of $a$ between $2$ and $10$ to within an error of $7\%$.

Notably, our best fit theoretical cumulative distribution function (CDF) does not reproduce the empirical CDF for $\tapp \gtrsim 0.2$.  This is not surprising giving the simplicity of our one parameter model, and that large apparent opening angle sources are more likely to have smaller viewing angles $\theta$.  Sources with small $\theta$ are more subject to differential Doppler beaming, where different streamlines in the jet have significantly different Doppler factors.  Other effects our simple model does not take into account, and that might explain the deviation between our data and best fit, are discussed below in the conclusion.

\section{Conclusion}
\label{conclusion}
We have illustrated a new method that measures the characteristic value of $\Gamma\theta_j$ in a flux-limited sample of jets, yielding $\Gamma\theta_j\approx 0.1 \pm 0.03$. This measurement implies that AGN jets are causally connected and therefore subject to a variety of processes requiring causal connection, such as large-scale magnetic reconnection and the development of global instabilities including the current-driven kink mode, pressure-driven modes, and Kelvin-Helmholtz modes.  

Some caution is needed in interpreting these results due both to our model's simplicity and the deviation of our best fit theoretical CDF from the empirical CDF for large apparent opening angle sources.  The derived PDF used to fit the empirical distribution of apparent opening angles crucially depends on the geometry of the emitting region, which we naively assumed to be a simple cone.  This is likely not the case in jets subject to relativistic velocity shear, where differential Doppler beaming causes different parts of the jet to become observable depending on the jet viewing angle.  Thus, the equation for how apparent opening angle depends on jet viewing angle and intrinsic opening angle (eqn. \ref{tri}) may need to be modified to reflect the effect of velocity shear.  Also, the assumptions implicit in Vermeulen \& Cohen's \cite{Vermeulen1994} calculation of the viewing angle PDF are open to criticism.  Namely, the result that viewing angle PDF only depends on the slope of the log-log integral source count plot, and not other parameters of the AGN jet luminosity function, needs to be further investigated.  We intend to address the above issues in a future publication.  Despite the above issues, it is encouraging that our results of $\Gamma\theta_j\approx0.1 \pm 0.03$ so closely aligns with previous measurements of $\Gamma\theta_j=0.13$ \cite{Pushkarev2009} and $\Gamma\theta_j=0.17\pm0.08$ \cite{Jorstad2005}, which used very different methods and assumptions.

\bibliographystyle{JHEP2}
\providecommand{\href}[2]{#2}\begingroup\raggedright\endgroup

\label{lastpage}

\end{document}